# Nematic spin fluid in the tetragonal phase of BaFe$_2$As$_2$

Leland W. Harriger [1], Huiqian Luo [2], Mengshu Liu [1], T. G. Perring [3], C. Frost [3], Jiangping Hu [4,2], M. R. Norman [5], and Pengcheng Dai [1,6,2]

[1] Department of Physics and Astronomy, The University of Tennessee, Knoxville, Tennessee 37996-1200, USA

[2] Beijing National Laboratory for Condensed Matter Physics, Institute of Physics, Chinese Academy of Sciences, Beijing 100190, China

[3] ISIS Facility, Rutherford Appleton Laboratory, Chilton, Didcot, Oxfordshire OX11 0QX, United Kingdom

[4] Department of Physics, Purdue University, West Lafayette, Indiana 47907, USA

[5] Materials Science Division, Argonne National Laboratory, Argonne, Illinois 60439, USA

[6] Neutron Scattering Science Division, Oak Ridge National Laboratory, Oak Ridge, Tennessee 37831, USA

**Magnetic interactions are generally believed to play a key role in mediating electron pairing for superconductivity in iron arsenides[1-4]; yet their character is only partially understood[5-8]. Experimentally, the antiferromagnetic (AF) transition is always preceded by or coincident with a tetragonal to orthorhombic structural distortion[5,6]. Although it has been suggested that this lattice distortion is driven by an electronic nematic phase[9-11], where a spontaneously generated electronic liquid crystal state breaks the C$_4$ rotational symmetry of the paramagnetic state[12], experimental evidence for electronic anisotropy has been either in the low-**

**temperature orthorhombic phase[7,13] or the tetragonal phase under uniaxial pressure that breaks this symmetry[14-16]. Here we use inelastic neutron scattering to demonstrate the presence of a large in-plane spin anisotropy above $T_N$ in the unstressed tetragonal phase of $BaFe_2As_2$. In the low-temperature orthorhombic phase, we find highly anisotropic spin waves[7] with a large damping along the AF $a$-axis direction. On warming the system to the paramagnetic tetragonal phase, the low-energy spin waves evolve into quasi-elastic excitations, while the anisotropic spin excitations near the zone boundary persist. These results strongly suggest that the spin nematicity we find in the tetragonal phase of $BaFe_2As_2$ is the source of the electronic and orbital anisotropy observed above $T_N$ by other probes[14-16], and has profound consequences for the physics of these materials.**

Correlated electron materials can exhibit a variety of complex phases that control the electronic and transport properties of these materials. For example, an electronic nematic phase, where the $C_4$ symmetry of the paramagnetic phase is spontaneously broken, has been postulated as the source of the pseudogap behavior observed in copper oxide superconductors[17-19]. Furthermore, the tetragonal to orthorhombic structural phase transition preceding or coincidental with the static AF order (with spin structure shown in Fig. 1a) in the parent compounds of iron pnictide superconductors[5,6] has been suggested to arise from a spin nematic phase[10,11]. Although neutron scattering[7], scanning tunneling microscopy[13], transport[14], optical conductivity[15], and angle resolved photoemission[16] experiments have provided evidence for electronic anisotropy, these measurements are carried out either in the low temperature orthorhombic phase[7,13], where the crystal lattice structure has already broken $C_4$ symmetry[5,6], or in the tetragonal phase under uniaxial pressure that also breaks this symmetry[14-16]. Therefore, it is unclear whether electronic anisotropy can exist in a truly tetragonal phase without an external driving field. A decisive answer to this question will not only reveal the microscopic origin of the lattice and magnetic transitions in iron arsenides, but will also determine the importance of electron correlations and orbital degrees of freedom in these materials[20-26].

Using inelastic neutron scattering, we first show that the spin waves of $BaFe_2As_2$ in the AF orthorhombic phase imply highly anisotropic magnetic exchange couplings similar to those seen in $CaFe_2As_2$ (Ref. 7) and these spin waves are strongly damped along the AF $a$-axis direction (Figs. 1c-f). This is consistent with transport measurements where the AF $a$-axis direction is more metallic than the ferromagnetic $b$-axis direction[14]. Upon warming the material to the tetragonal paramagnetic phase[5,6], the anisotropic high-energy (>100 meV) spin excitations near the zone boundary persist, while the low-energy spin waves near the zone center evolve into paramagnetic spin excitations (Figs. 2-4). These results provide compelling evidence for a nematic spin fluid that breaks the tetragonal $C_4$ symmetry of the underlying crystalline lattice and spontaneously forms without the need for uniaxial pressure. Moreover, we suggest that this spin anisotropy causes a splitting of the $d_{xz}$ and $d_{yz}$ orbital bands in the tetragonal phase[20-26], which in turn leads to the orthorhombic lattice distortion and electronic anisotropy.

We use inelastic neutron scattering to study the temperature dependent spin waves of single crystals of $BaFe_2As_2$ with a Néel temperature of $T_N \approx 138$ K (Ref. 6). Previous powder[27] and single crystal[28] measurements for excitation energies below 100 meV revealed that the spin waves in $BaFe_2As_2$ are three-dimensional and centered at the AF wave vector $Q = (1, 0, L = 1, 3, 5, …)$ in reciprocal lattice units (rlu). For $CaFe_2As_2$, spin waves form well-defined ellipses centered around the AF wave vector $Q$ throughout the Brillouin zone[7]. Figs. 2a-e show two-dimensional constant-energy ($E$) images of spin-wave excitations of $BaFe_2As_2$ in the ($H$, $K$) scattering plane for several Brillouin zones at $L = 1, 3, 5, 7$. For energy transfers of $E = 26 \pm 10$ (Fig. 2a) and $81 \pm 10$ meV (Fig. 2b), spin waves are still peaked at $Q = (1,0)$ in the center of the Brillouin zone shown as dashed square boxes. As the energy increases to $E = 113 \pm 10$ (Fig. 2c), $157 \pm 10$ (Fig. 2d), and $214 \pm 15$ meV (Fig. 2e), spin waves no longer form ellipses centered around $Q = (1,0)$. Instead, they start to split along the $K$-direction and form an anisotropic and asymmetric ring around $Q = (\pm1, \pm1)$, in stark contrast with the spin waves at similar energies seen in $CaFe_2As_2$ (Figs. 1e-1i of Ref. 7).

To understand the low-temperature spin waves in $BaFe_2As_2$, we cut through the two-dimensional images similar to Fig. 2 for incident beam wave vectors ($k_i$) aligned along the $c$-axis. Figs. 1e and 1f show spin wave dispersions along the $(1,K)$ and $(H,0)$ directions, respectively. Fig. 2f shows the background subtracted scattering for the $E_i = 450$ meV data projected in the wave vector ($Q = [1,K]$) and energy space. Similar to spin waves in $CaFe_2As_2$ (Ref. 7), we can see three clear plumes of scattering arising from the in-plane AF zone centers $Q = (1,-2), (1,0)$, and $(1,2)$ extending up to about 200 meV. We have attempted but failed to fit the entire spin wave spectra in Fig. 2 using a Heisenberg Hamiltonian consisting of effective in-plane nearest-neighbors (Fig. 1a, $J_{1a}$ and $J_{1b}$), next-nearest-neighbor (Fig. 1a, $J_2$), and out-of-plane ($J_c$) exchange interactions with an isotropic spin wave damping parameter $\Gamma$ (black curves in Fig. 1c and supplementary information)[7]. However, allowing for an anisotropic spin wave damping parameter $\Gamma$ (Fig. 1d) produces an energy dependence of the spin wave profiles (color plots in Fig. 1c) that is qualitatively similar to what we observe (Figs. 2a-2e). Using this wave vector dependent damping $\Gamma(H, K)$ (see supplementary information), we were able to fit the entire measured spin wave excitation spectra in absolute units by convolving the neutron scattering spin-wave cross section with the instrument resolution[7]. The effect of twin domains is taken into account by *a/b* averaging (see supplementary information). Consistent with earlier results on CaFe$_2$As$_2$ (Ref. 7), we find that the Heisenberg Hamiltonian with $SJ_{1a} \approx SJ_{1b} \approx \frac{1}{2}SJ_2$ fails to describe the zone boundary data (Fig. 1e). Our best fits to both the low-energy and zone boundary spin waves by independently varying the effective exchange parameters are shown as solid lines in Fig. 1e and color plots in Figs. 2g-l with $SJ_{1a} = 59.2 \pm 2.0$, $SJ_{1b} = -9.2 \pm 1.2$, $SJ_2 = 13.6 \pm 1.0$, $SJ_c = 1.8 \pm 0.3$ meV.

Comparing the above fitted results for BaFe$_2$As$_2$ with those for CaFe$_2$As$_2$ (Ref. 7), we see that while the in-plane effective magnetic exchanges $(SJ_{1a}, SJ_{1b})$ are very similar in these two materials, there is ~30% reduction in $SJ_2$ when Ca is replaced by the larger Ba and the *c*-axis exchange coupling is reduced considerably (from $SJ_c = 5.3 \pm 1.3$ meV for CaFe$_2$As$_2$). In addition, while one can see clear spin wave ellipses centered around $Q$ = (1,0) in CaFe$_2$As$_2$ at all energies[7], spin waves in BaFe$_2$As$_2$ are heavily damped along the *a*-axis direction and become hardly observable for energies above 100 meV (see supplementary information), suggesting that the spin waves cross the Stoner continuum along the *a*-direction.

Having demonstrated that BaFe$_2$As$_2$ exhibits a large spin anisotropy in the low-temperature orthorhombic phase, it is important to determine if this spin anisotropy also exists in the high-temperature tetragonal phase, where the underlying crystal lattice structure has $C_4$ rotational symmetry. In a recent inelastic neutron scattering study on CaFe$_2$As$_2$, spin excitations in the paramagnetic tetragonal phase were found to have a similar spatial line-shape as those of the low-temperature spin waves below 60 meV (Ref. 29). These anisotropic short-range AF fluctuations can be interpreted as frustrated paramagnetic scattering[29]. If the observed large anisotropy of $SJ_{1a}$ and $SJ_{1b}$ for BaFe$_2$As$_2$ (Figs. 1 and 2) and CaFe$_2$As$_2$ (Ref. 7) in the AF orthorhombic phase becomes isotropic ($SJ_{1a} = SJ_{1b}$) in the paramagnetic tetragonal phase, one would expect a huge softening of the zone boundary spin waves upon entering into the tetragonal phase (see dotted lines in Fig. 1e), which we do not observe. Figure 3 summarizes the temperature dependence of the spin wave excitations at temperatures of $0.05T_N$, $0.93T_N$, and $1.09T_N$. For spin wave energies of $E = 50 \pm 10, 75 \pm 10$ meV, we confirm the earlier result[29] on CaFe$_2$As$_2$ and find that spin excitations above $T_N$ are weaker and broader than the spin waves below $T_N$ (Figs. 3a-f). However, we discovered that spin waves at energies of $E = 125 \pm 10, 150 \pm 10$ meV have virtually no temperature dependence of their intensity and line shape across the AF orthorhombic to paramagnetic tetragonal phase transition (Figs. 3g-l). Therefore, spin excitations near the zone boundary do not exhibit huge softening in the paramagnetic state, which implies that the large in-plane exchange anisotropy persists above $T_N$ without spin frustration.

To test whether the observed scattering above $T_N$ indeed arises from localized spin excitations similar to the spin waves below $T_N$ and not from paramagnetic scattering centered at zero energy, we carried out energy cuts of the spin excitations at different positions of the dispersion as shown in the inset of Fig. 4a. Near the Brillouin zone center at $Q = (1,0.05)$ and $(1,0.02)$, well-defined spin waves are observed at $E = 32$, and

50 meV, respectively (blue diamonds in Figs. 4a and 4b), in the AF ordered state. Upon warming to the paramagnetic tetragonal state ($T = 1.09T_N$), the spin wave peaks disappear and spin excitations become purely paramagnetic with their highest intensity centered at zero energy (red circles in Figs. 4a and 4b). Moving closer to the zone boundary at $Q = (1,0.35)$, the spin wave peaks at 90 meV are virtually unchanged on warming from $0.05T_N$ to $0.93T_N$ and decrease only slightly in intensity at $1.09T_N$ (Fig. 4c). At $Q = (1,0.5)$, spin wave peaks at $E = 125$ meV are temperature independent below and above $T_N$ (Fig. 4d). Figs. 4e and 4f show the wave vector dependence of the magnetic scattering at $E = 19 \pm 5$ and $128 \pm 5$ meV, respectively. Consistent with the results in Fig. 3, the spin waves at low energies become broad paramagnetic spin excitations above $T_N$, while they stay unchanged at high energies near the zone boundary (Figs. 4e and 4f). The energy dependence of the dynamic spin-spin correlation lengths below and above $T_N$ in Fig. 4g suggests that short-range spin excitations at energies above ~100 meV are not sensitive to the orthorhombic to tetragonal phase transition and do not reflect the $C_4$ symmetry. The effective magnetic exchange couplings $SJ_{1a}$ and $SJ_{1b}$ in spin clusters of sizes $\xi = 15 \pm 3$ Å must be anisotropic, and therefore locally break the $C_4$ tetragonal symmetry.

We have discovered that the spin waves in BaFe$_2$As$_2$ are highly anisotropic with a large damping along the metallic AF $a$-axis direction in the AF orthorhombic phase (Figs. 1 and 2). On warming the material to the paramagnetic tetragonal phase, the low-energy spin waves near the zone center evolve into paramagnetic scattering, while the anisotropy of the high-energy spin excitations near the zone boundary persists (Figs. 3 and 4). This means that the short-range effective magnetic exchange couplings in BaFe$_2$As$_2$ are anisotropic and unchanged across $T_N$, consistent with a nematic spin fluid that breaks the $C_4$ symmetry of the tetragonal phase. In previous observations of electronic nematic phases in different materials, there is usually a symmetry breaking field present, such as an external magnetic field, uniaxial pressure, or an orthorhombic crystalline lattice[12-19], which is not the case here. The observation of a short-range spin nematic phase in the tetragonal state of BaFe$_2$As$_2$ reveals the presence of strong spin-orbital coupling at temperatures above $T_N$ (Refs. 17, 23-26, 30).

The persistence of spin anisotropy in the paramagnetic phase has obvious implications for the nature of the magnetism in pnictides, which in turn has potentially profound implications for the origin of superconductivity. Anisotropy in the resistivity has been seen to persist for Co doped BaFe$_2$As$_2$ samples into the region of the phase diagram where superconductivity exists[14]. Moreover, the existence of a spin resonance in the superconducting state of Ni doped BaFe$_2$As$_2$, which is a doublet rather than a triplet, is also consistent with local spin nematicity[31]. Since the spin excitations at short length scales are intrinsically nematic in the paramagnetic tetragonal phase, the AF phase transition and lattice distortion are likely induced by nematic spin fluctuations. On the other hand, if orbital ordering were driving the spin nematicity, one would expect a gradual change of spin anisotropy across $T_N$ depending on the strength of spin-orbital coupling, contrary to our observations. Since the spin nematicity leads to an enormous anisotropy in the near-neighbor exchange couplings, this could have a profound impact on the nature of the superconducting electron pairing interaction. In that connection, it is interesting to note that there appears to be an anti-correlation between the spin nematicity and the superconducting gap anisotropy, in that the latter appears to switch from *s*-wave-like to *d*-wave-like[32] at a doping where the spin nematicity disappears in the transport measurements[14].

**Acknowledgements** This work is supported by the US NSF and DOE Division of Materials Science and Engineering, Basic Energy Sciences.  This work is also supported by the US DOE through UT/Battelle LLC.  The work at IOP is supported by the Chinese Academy of Sciences.

**Author Information**  The authors declare no competing financial interests. Correspondence and requests for materials should be addressed to P.D. (daip@ornl.gov).

**Figure 1 Magnetic structure, transport measurements, measured/calculated spin-wave dispersions, and wave vector dependence of damping anisotropy for $BaFe_2As_2$.** Our inelastic neutron scattering experiments were carried out on the MAPS time-of-flight chopper spectrometer at the Rutherford-Appleton Laboratory, Didcot, UK. We co-aligned ~25 grams of single crystals of $BaFe_2As_2$ grown by self-flux (with in-plane mosaic of 2 degrees and out-of-plane mosaic of 3 degrees). The incident beam energies were $E_i$ = 80, 200, 250, 450, 600 meV, and with $k_i$ parallel to the *c*-axis. Spin wave intensities were normalized to absolute units using a vanadium standard (with 30% error). We define the wave vector $Q$ at $(q_x,q_y,q_z)$ as $(H, K, L) = (q_x\, a/2\pi, q_y\, b/2\pi, q_z\, c/2\pi)$ in rlu, where $a$ = 5.62, $b$ = 5.570, and $c$ = 12.97 Å are the orthorhombic cell lattice parameters at 10 K (Ref. 6). a) Schematic diagram of the Fe spin ordering in $BaFe_2As_2$ with the effective magnetic exchange couplings $J_{1a}, J_{1b}, J_2$ along different directions. b) Temperature dependence of the resistivity in detwinned $BaFe_2As_2$ (from Ref. 15). The inset is a plot of the resistivity for the twinned sample used in our neutron measurements, with the blue points corresponding to $T = 7, 125, 150$ K. c) Color plots describing qualitatively how the spin wave scattering evolves from $Q$ = (1, 0) to (1, 1) as a function of energy. The plots are constructed using the Heisenberg model described in the text with an anisotropic damping $\Gamma$. The solid black contours are an overlay of the same model but with no damping. The exchange couplings used in both plots are from best fits of the data; however the damping parameters are floated to values that are most representative of how the spin waves evolve as a function of energy. d) Color plots of the anisotropic damping $\Gamma$, which is much stronger along the *H*-direction than along the *K*-direction. e) Spin wave dispersion along the (1, *K*) direction as determined by energy and

$Q$-cuts of the raw data in Fig. 2 below and above $T_N$. The solid line is a Heisenberg model calculation using anisotropic exchange couplings $SJ_{1a} = 59.2 \pm 2.0$, $SJ_{1b} = -9.2 \pm 1.2$, $SJ_2 = 13.6 \pm 1.0$, $SJ_c = 1.8 \pm 0.3$ meV determined by fitting the full cross-section. The dotted line is a Heisenberg model calculation assuming isotropic exchange coupling $SJ_{1a} = SJ_{1b} = 18.3 \pm 1.4$, $SJ_2 = 28.7 \pm 0.5$, and $SJ_c = 1.8$ meV. f) Dispersion along the $(H, 0)$ direction: data points beyond $H = 1.4$ could not be reliably obtained due to strong damping at higher energies. The red shading stresses how the damping grows as a function of $H$. Error bars are systematic and represent the difference between $Q$ and $E$ cut dispersion points. The statistical error of the $Q$ and $E$ cuts are much smaller.

**Figure 2 Images of the spin waves as a function of increasing energy at $T = 7$ K and our model fit using the Heisenberg Hamiltonian.** Wave vector dependence of the spin waves for energy transfers of a) $E = 26 \pm 10$ meV [$E_i = 450$ meV and $Q = (H, K, 1)$]; b) $E = 81 \pm 10$ meV [$E_i = 450$ meV and $Q = (H, K, 3)$]; c) $E = 113 \pm 10$ meV [$E_i = 450$ meV and $Q = (H, K, 5)$]; d) $E = 157 \pm 10$ meV [$E_i = 600$ meV and $Q = (H, K, 5)$]; e) $E = 214 \pm 10$ meV [$E_i = 600$ meV and $Q = (H, K, 7)$]; f) The projection of the spin waves on the energy transfer axis and $(1, K)$ direction (with integration of $H$ from 0.8 to 1.2 rlu) after subtracting the background integrated from $1.8 < H < 2.2$ and from $-0.25 < K < 0.25$ with $E_i = 450$ meV. The color bar scales represent the vanadium-normalized absolute spin wave intensity in units of mbarn/sr/meV/f.u and the dashed boxes indicate zone boundaries. g-l) Model calculation of identical slices as in

a-f) using anisotropic exchange couplings from best fits and convolved with the instrumental resolution.

**Figure 3 Temperature dependence of the spin waves at different energies across the AF orthorhombic to paramagnetic tetragonal phase transition.** Spin waves of $E = 50 \pm 10$ meV a-c); $E = 75 \pm 10$ meV d-f); $E = 125 \pm 10$ meV g-i); and $E = 150 \pm 10$ meV j-l) for temperatures of $T = 7$, 125, and 150 K. The color bars represent absolute spin wave intensity in units of mbarn/sr/meV/f.u., and dashed curves show fixed reciprocal space sizes at different temperatures.

**Figure 4 Energy(*E*)/wave vector (*Q*) dependence of the spin wave excitations and dynamic spin-spin correlation lengths (ξ) at different energies below and above $T_N$.** The blue diamonds in a-d) are constant-*Q* cuts at ***Q*** = (1,0.05), (1,0.2), (1,0.35), and (1,0.5), respectively, at $T = 7$ K. The green squares and red circles in a-d) are identical constant-*Q* cuts at $T = 125$, and 150 K, respectively. The dashed lines indicate paramagnetic scattering at low energies centered at $E = 0$ meV. Above ~100 meV, the high temperature spin excitations follow the same dispersion and intensity as the spin wave data below $T_N$. e,f) Wave vector dependence of the spin wave excitations below and above $T_N$ obtained through constant-*E* cuts at $E = 19 \pm 5$ and $128 \pm 5$ meV. The solid lines in a-f) are fits to the anisotropic spin-wave model discussed in the text and the horizontal bars represent the instrumental energy (*E*)/wave vector (*Q*) resolution. g) Energy dependence of the dynamic spin-spin correlation lengths below and above $T_N$ obtained by Fourier transform of constant-*E* cuts similar to e,f). For excitation energies

above 100 meV, the dynamic spin-spin correlation lengths are independent of the AF orthorhombic to paramagnetic tetragonal phase transition.

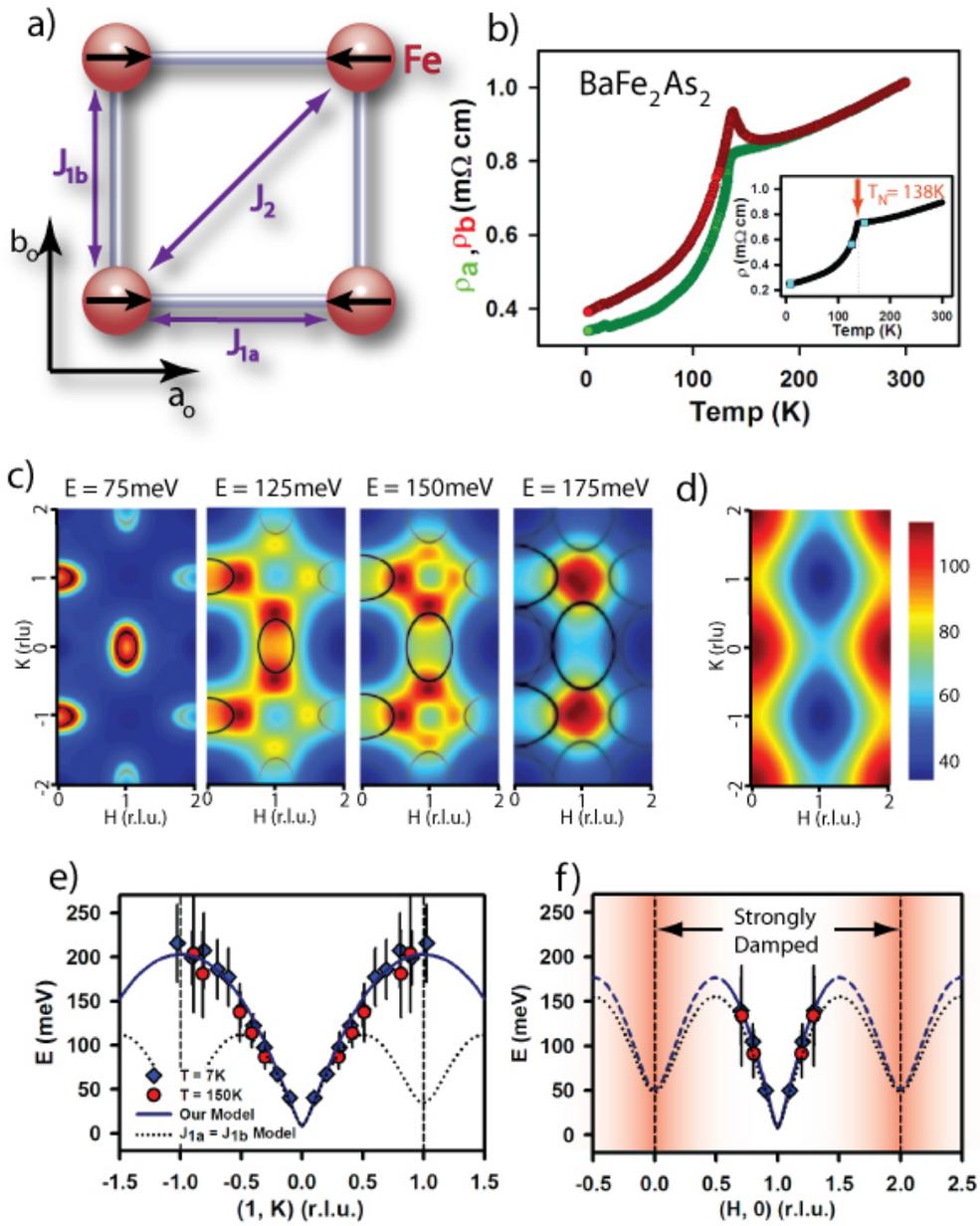

Fig. 1

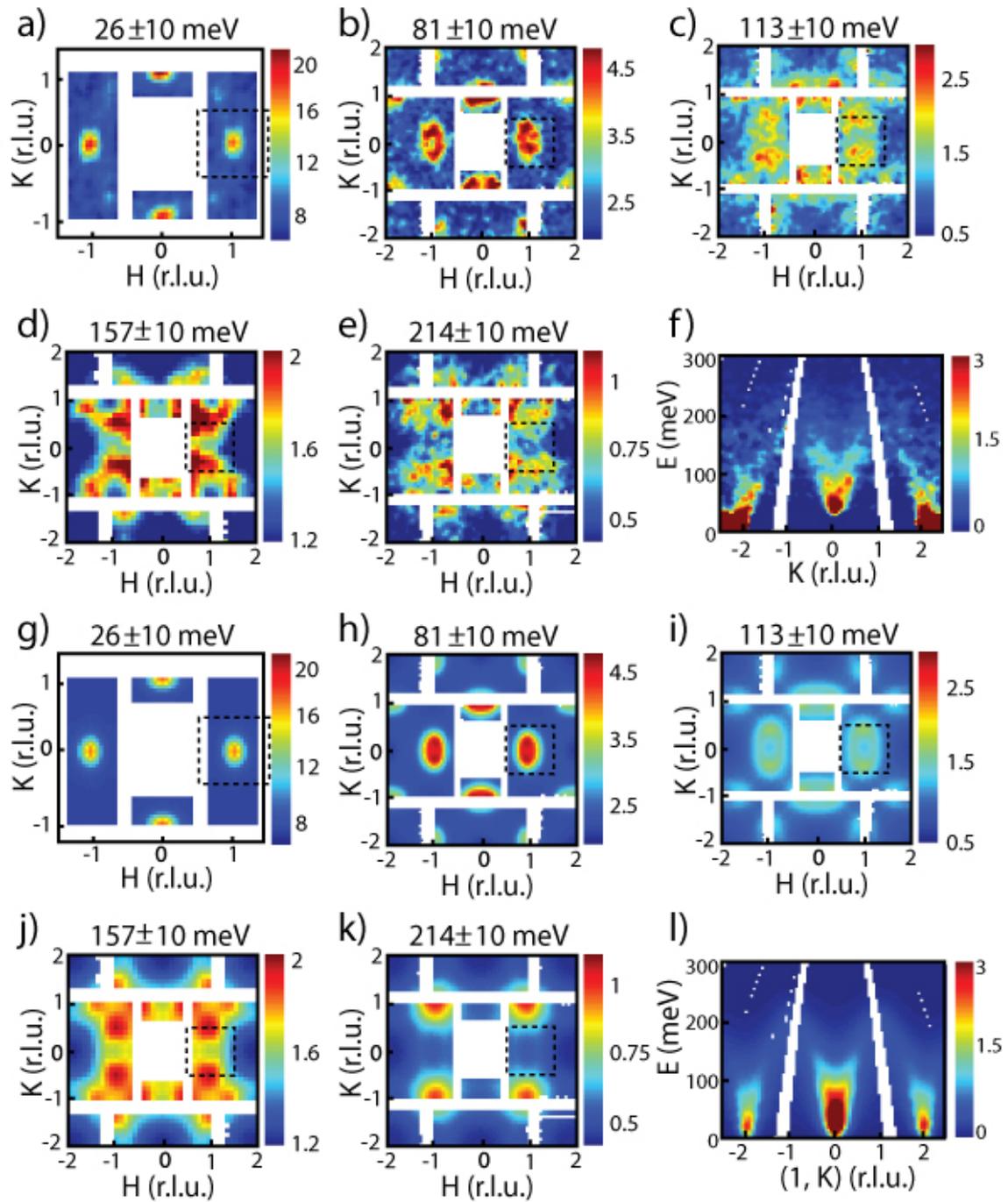

Fig. 2

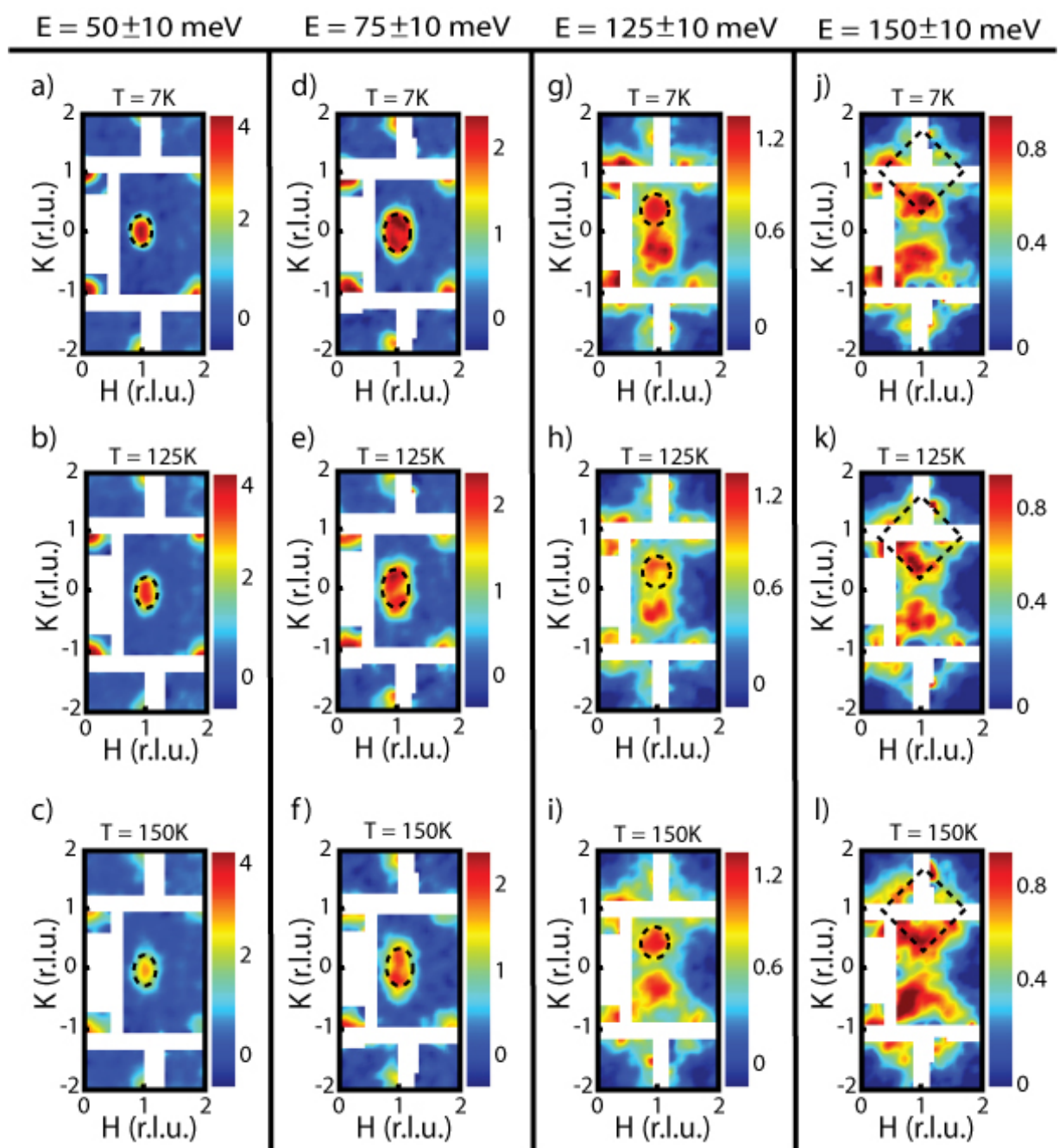

Fig. 3

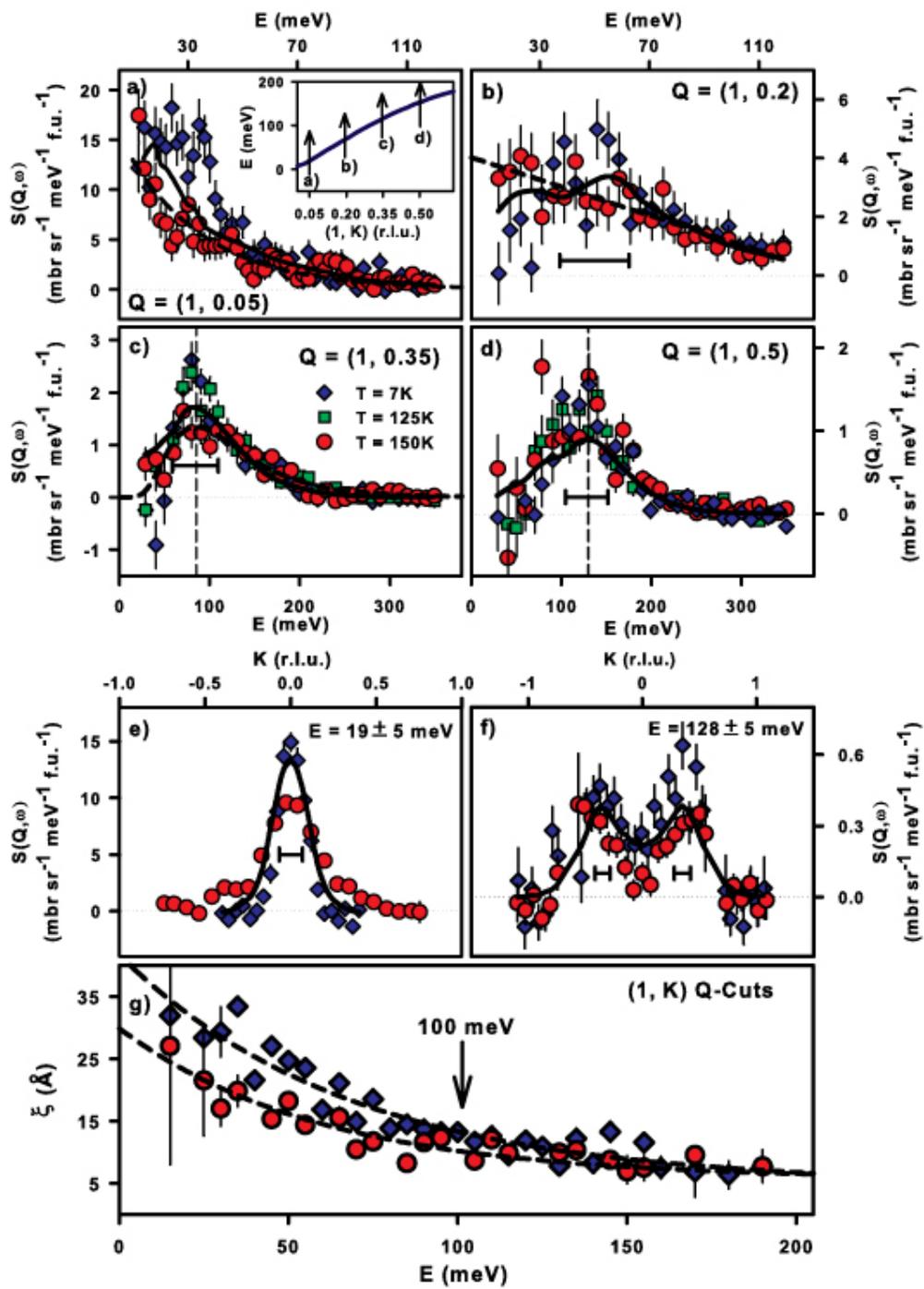

Fig. 4

## Supplementary information:

To understand the spin wave data as shown in Figs. 1-4, we consider a Heisenberg Hamiltonian consisting of effective in-plane nearest-neighbors (Fig. 1a, $J_{1a}$ and $J_{1b}$), next-nearest-neighbor (Fig. 1a, $J_2$), and out-of-plane ($J_c$) exchange interactions. The dispersion relations are given by: $E(q) = \sqrt{A_q^2 - B_q^2}$, where $A_q = 2S[J_{1b}(\cos(\pi K) - 1) + J_{1a} + J_c + 2J_2 + J_s]$,

$B_q = 2S[J_{1a}\cos(\pi H) + 2J_2\cos(\pi H)\cos(\pi K) + J_c\cos(\pi L)]$, $J_s$ is the single ion anisotropy constant, and $q$ is the reduced wave vector away from the AF zone center. The neutron scattering cross section can be written as:

$$\frac{d^2\sigma}{d\Omega dE} = \frac{k_f}{k_i}\left(\frac{\gamma r_0}{2}\right)^2 g^2 f^2(Q) e^{-2W} \sum_{\alpha\beta}(\delta_{\alpha\beta} - Q_\alpha Q_\beta) S^{\alpha\beta}(Q,E),$$ where

$(\gamma r_0/2)^2 = 72.65$ mb/sr, $g$ is the $g$-factor ($\approx 2$), $f(Q)$ is the magnetic form factor of iron $Fe^{2+}$, $e^{-2W}$ is the Debye-Waller factor ($\approx 1$ at 10 K), $Q_\alpha$ is the $\alpha$ component of a unit vector in the direction of $Q$, $S^{\alpha\beta}(Q,E)$ is the response function that describes the $\alpha\beta$ spin-spin correlations, and $k_i$ and $k_f$ are incident and final wave vectors, respectively. Assuming that only the transverse correlations contribute to the spin-wave cross section and finite excitation lifetimes can be described by a damped simple harmonic oscillator with inverse lifetime $\Gamma$, we have

$$S^{yy}(Q,E) = S^{zz}(Q,E) = S_{eff} \frac{(A_q - B_q)}{E_0(1 - e^{-E/k_B T})} \frac{4}{\pi} \frac{\Gamma E E_0}{(E^2 - E_0^2)^2 + 4(\Gamma E)^2},$$

where $k_B$ is the Boltzmann constant, $E_0$ is the spin-wave energy, and $S_{eff}$ is the effective spin. Assuming isotropic spin wave inverse lifetime $\Gamma$, we were unable to find any effective exchange couplings that will describe the entire spin wave spectra as shown in Fig. 2a-f. To resolve this problem, we have used an anisotropic spin wave damping $\Gamma$ assuming $\Gamma(H,K) = \Gamma_0 + \Gamma_1 E + A(\cos(\frac{\pi H}{2}))^2 + B(\cos(\frac{\pi K}{2}))^2$, where $A$ and $B$ are parameters controlling the magnitude of the spin wave damping. For the best fit to the spin wave data, we have $\Gamma_0 = 32 \pm 10.6, \Gamma_1 \rightarrow 0, A = 51.9 \pm 9.0, B = 27.8 \pm 7.3$ with magnetic exchange couplings as listed in the main text.

To illustrate how neutron scattering can probe spin waves in two high symmetry directions of twinned samples, we note that in the AF orthorhombic phase, the static AF order occurs at the AF wave vector $Q$ = (1, 0, $L$ = 1,3,5…) rlu and the AF Bragg peak is NOT allowed at $Q$ = (0, 1, $L$ = 1,3,5…) rlu (Ref. S1). Therefore, spin waves originating from each of the twin domains of the BaFe$_2$As$_2$ in the AF orthorhombic phase will not overlap until they are near the zone boundary. Figure SI1 shows spin wave intensity calculations as a function of energy for twinned and detwinned BaFe$_2$As$_2$ using identical parameters as discussed in the text. For most spin wave energies of interest, the effect of twinning is simply to have two single domain excitations rotated by 90 degrees (Figs. SI1a-d).

Figure SI2 shows our calculated dispersion curves in the case of twinned and single domain samples. As one can see from the spectra, the effect of twinning will only become important near the top of the band with a very small intensity contribution. Figure SI3 shows constant energy cuts of the spin wave dispersions along two high symmetry directions as a function of increasing energy and our model fit using the Heisenberg Hamiltonian with anisotropic damping as discussed in the text. The solid lines are model fits to the data after convolving the cross section with the instrumental

resolution. Both the intensity and linewidth of the excitations are considered in the model.

To demonstrate that the $J_{1a} = J_{1b}$ Heisenberg Hamiltonian cannot describe the high-energy zone boundary spin wave data, we show in Fig. SI4 the best fit of the low-energy spin wave data with $SJ_{1a} = SJ_{1b} = 18.3 \pm 1.4$, $SJ_2 = 28.7 \pm 0.5$, $SJ_c = 1.85$, $SJ_s = 0.084$ meV, and isotropic spin wave damping $\Gamma = 21 \pm 2$ (Ref. 7). We have calculated both the detwinned and twinned case. It is clear that the line-shape and intensity of the high-energy spin waves for this model disagree with the observation in Fig. 2. Fig. SI5 and Fig. SI6 show the output from the best fit of the $SJ_{1a} = SJ_{1b}$ model to the spin wave data. As one can see, the fit describes the low-energy spin wave data fairly well but fails to account for the high-energy zone boundary spin wave data.

Finally, to illustrate the dramatic difference in high-energy spin waves between $BaFe_2A_2$ and $CaFe_2As_2$, we show in Fig. SI7 constant-energy images of the spin waves for these two materials. Since the AF structure, twinning, and lattice structure of $BaFe_2As_2$ and $CaFe_2As_2$ are identical, one would expect that the effective AF exchange couplings in these materials should be similar. Inspection of Fig. SI7 reveals that spin waves of $BaFe_2A_2$ at $E = 144 \pm 15$ meV no longer form a ring centered around the AF ordering wave vector as in the case of $CaFe_2As_2$. The only way to interpret these data is to assume that spin waves along the (1,0) direction are heavily damped and no longer observable for $BaFe_2As_2$.

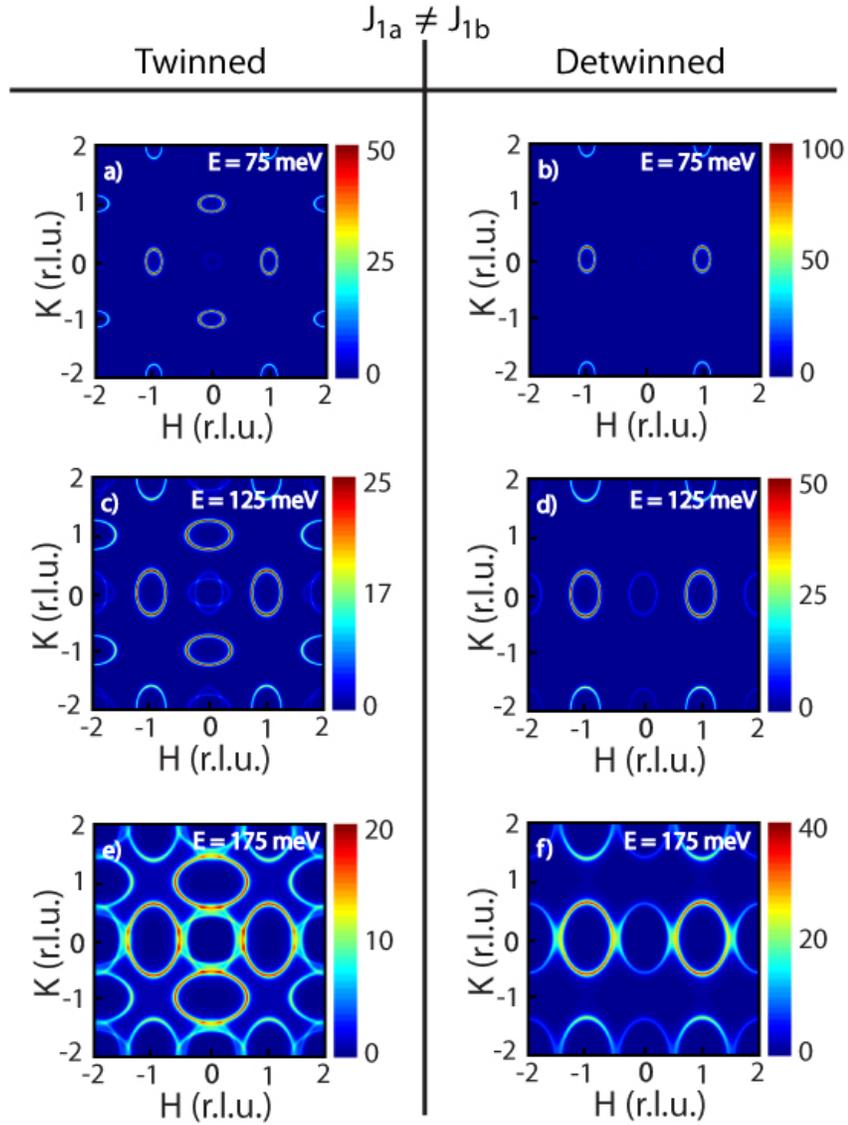

Fig. SI1 The effect of twinning on spin waves of BaFe$_2$As$_2$ with $L$ = odd.

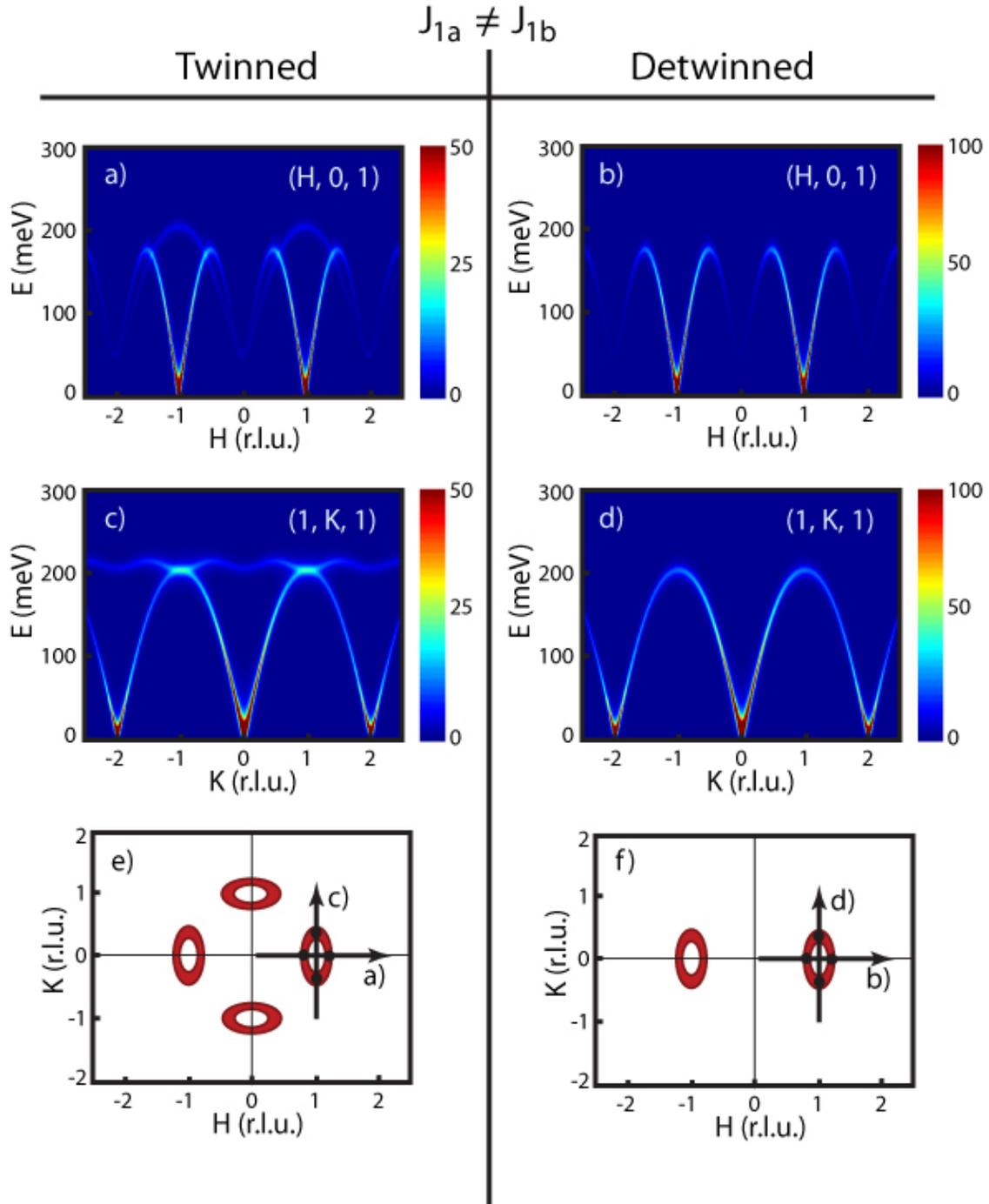

Fig. SI2 The effect of twinning on the dispersion curves of $BaFe_2As_2$.

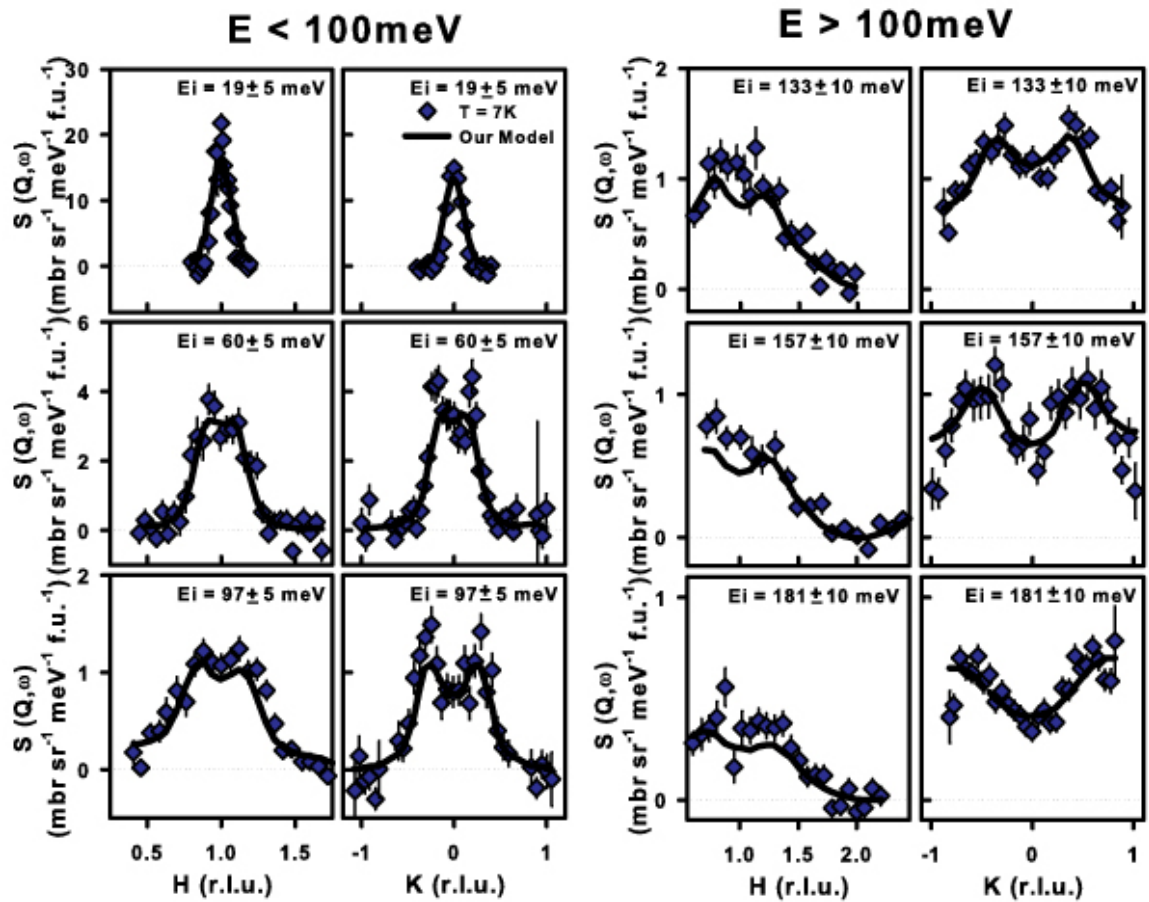

Fig. SI3 Constant energy cuts of the spin wave excitations at 7 K and our model fits to the data using an anisotropic Heisenberg Hamiltonian convolved with the instrumental resolution. The solid lines are the output from the Toby fit program (Ref. S2) using fitting parameters as discussed in the text of the paper.

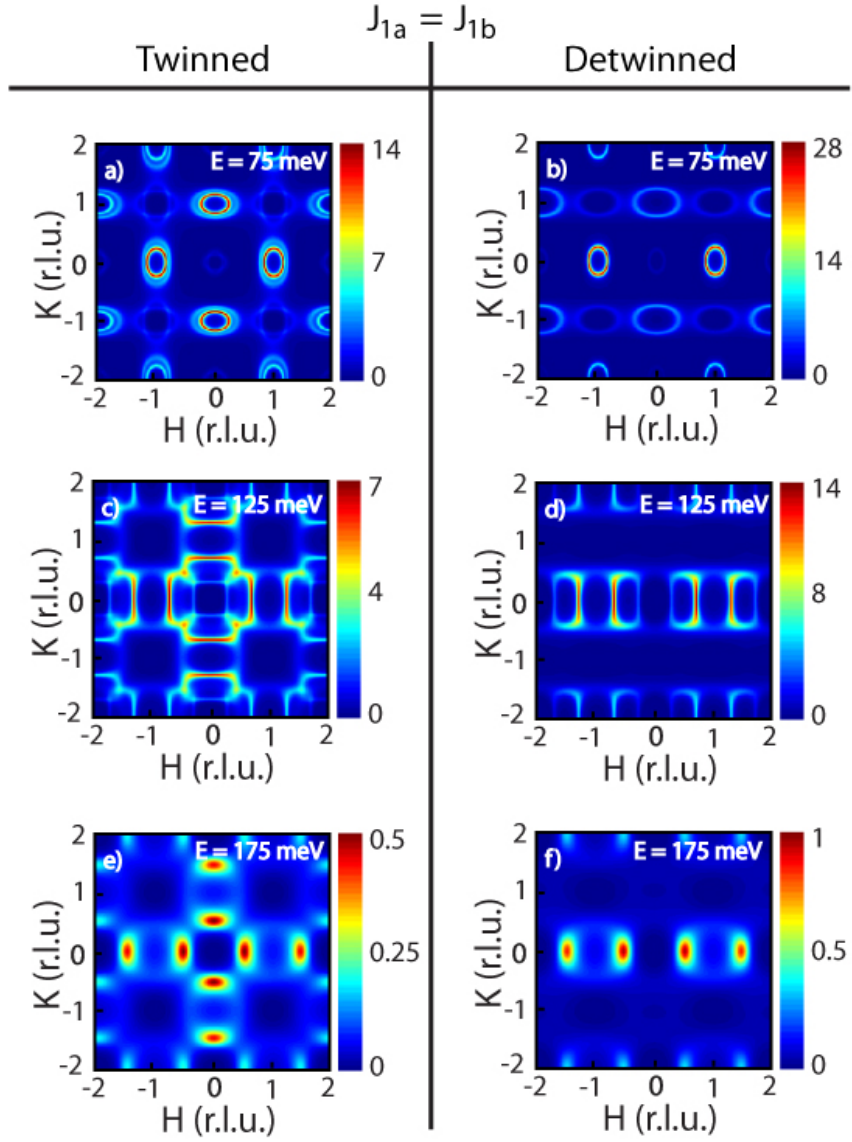

Fig. SI4 The effect of twinning on spin waves of BaFe$_2$As$_2$ with isotropic $J_{1a}$ and $J_{1b}$ and $L$ = odd.

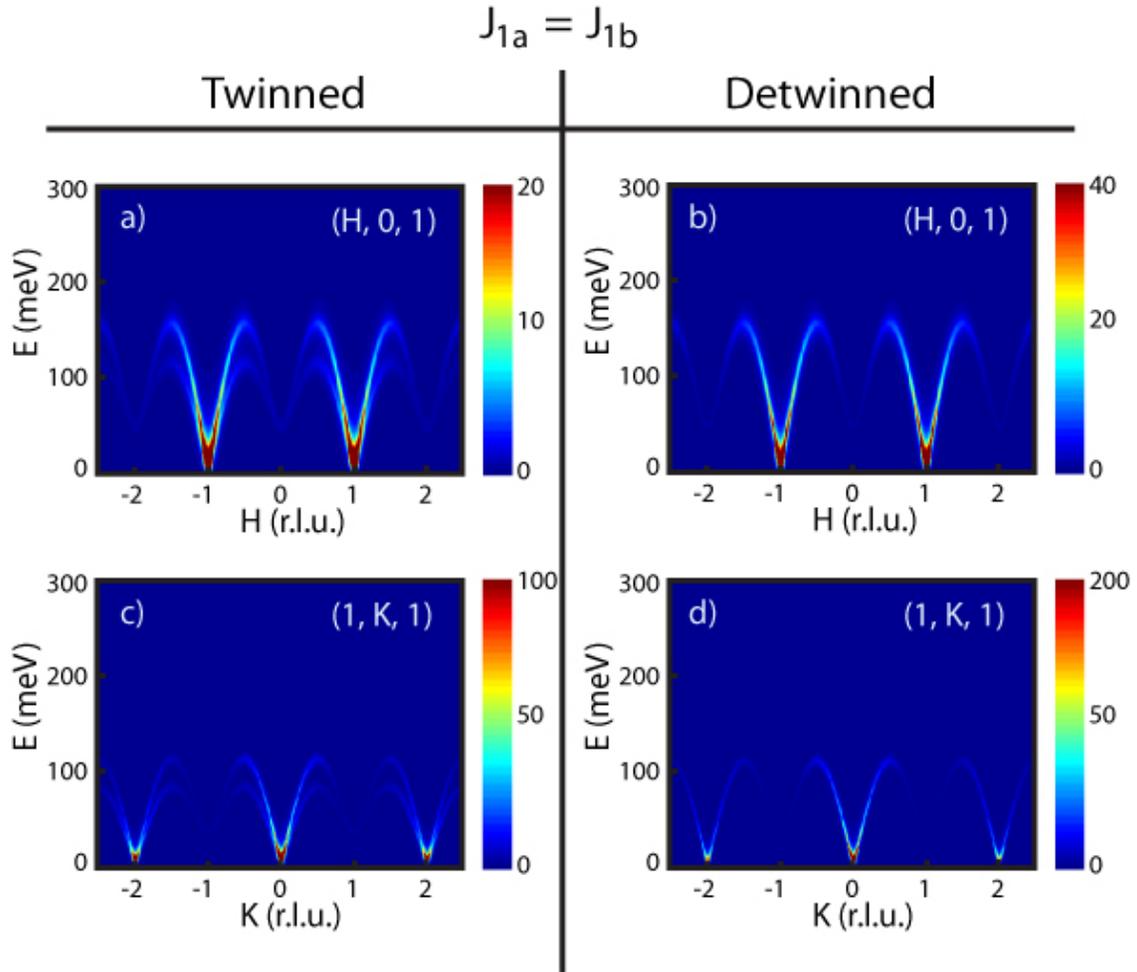

Fig. SI5 The effect of twinning on the dispersion curves of $BaFe_2As_2$ in the $J_{1a}$ and $J_{1b}$ model.

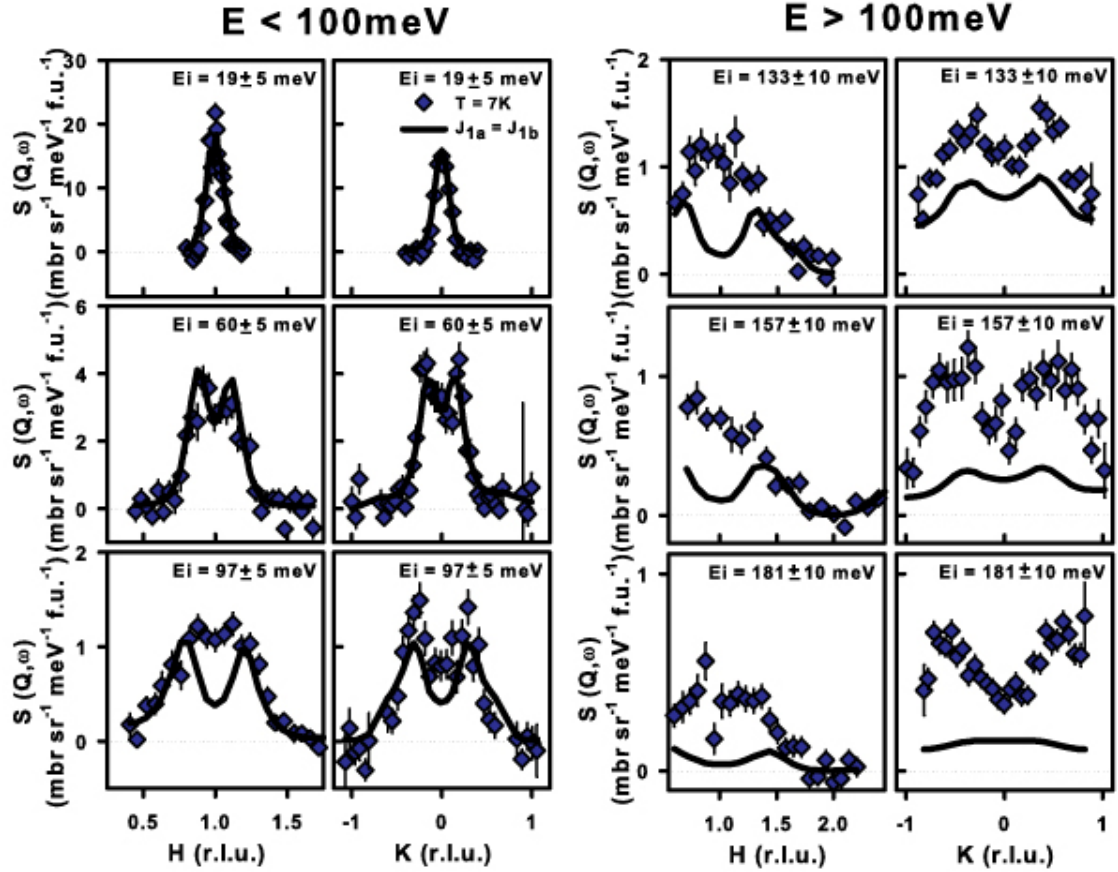

Fig. SI6 Constant energy cuts of the spin wave excitations at 7 K and the $J_{1a} = J_{1b}$ model fits to the data using an anisotropic Heisenberg Hamiltonian convolved with the instrumental resolution. The solid lines are the output from the Toby fit program (Ref. S2) using fitting parameters as discussed in the supplementary material. While this model fits the low-energy spin wave data reasonably well, it completely fails to describe the data for spin wave energies above 100 meV.

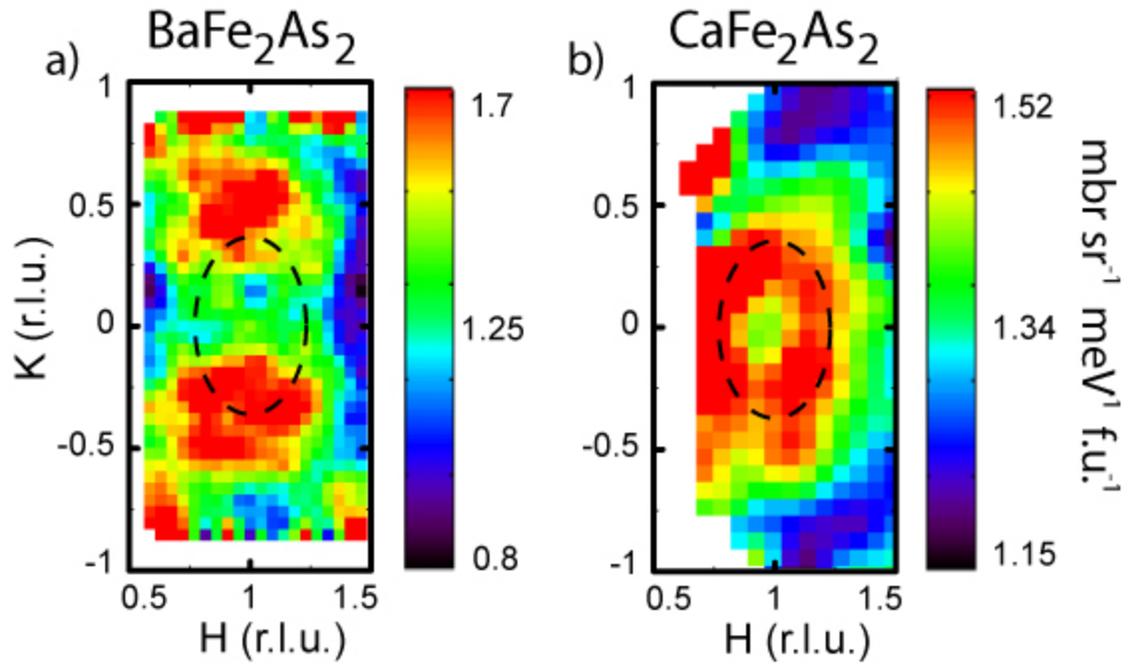

Fig. SI7 Constant energy cuts of the spin wave excitations at 7 K for $BaFe_2As_2$ and $CaFe_2As_2$ in absolute units within the first Brillouin zone. The data for $CaFe_2As_2$ are from Ref. 7.